\documentclass[journal,twoside,web]{ieeecolor}
\usepackage{jsen}
\usepackage{cite}
\usepackage{amsmath,amssymb,amsfonts}
\usepackage{algorithmic}
\usepackage{graphicx}
\usepackage{textcomp}
\usepackage{wrapfig}

\usepackage{setspace} % 行間隔を調整（表で使う場合/表の幅が文と同じになる）
\makeatletter
\def\subsection{%
\@startsection{subsection}{2}{\z@}%
{3.5ex plus 1.5ex minus 1.5ex}%
{0.7ex plus .5ex minus 0ex}%
{\leavevmode\color{subsectioncolor}\normalsize\normalfont\everymath={\sf}\sf\itshape\raggedright}%
}
\makeatother

\def\BibTeX{{\rm B\kern-.05em{\sc i\kern-.025em b}\kern-.08em
T\kern-.1667em\lower.7ex\hbox{E}\kern-.125emX}}
\definecolor{abstractbg}{rgb}{0.89804,0.94510,0.83137}
\begin{document}
\title{Self-Noise Reduction for Capacitive Sensors \\via Photoelectric DC Servo: \\Application to Condenser Microphones}
\author{Hirotaka Obo, Atsushi Tsuchiya, Tadashi Ebihara and Naoto Wakatsuki
% \thanks{This work was partly supported by JSPS KAKENHI Grant Numbers JP23H01617 and JP21J10818. } % not use
\thanks{Hirotaka Obo is with the Institute for Rural Engineering, National Agriculture and Food Research Organization (NARO), 2-1-6 Kannondai, Tsukuba, Ibaraki, Japan (e-mail: obo.hirotaka468@naro.go.jp). }
\thanks{Atsushi Tsuchiya, Ebihara Tadashi and Wakatsuki Naoto are with the Faculty of Engineering, Information and Systems, University of Tsukuba, 1-1-1 Tennodai, Tsukuba, Ibaraki, Japan (e-mail: tsuchiya@aclab.esys.tsukuba.ac.jp; ebihara@iit.tsukuba.ac.jp; wakatuki@iit.tsukuba.ac.jp).}
}
\IEEEtitleabstractindextext{%
\fcolorbox{abstractbg}{abstractbg}{%
\begin{minipage}{\textwidth}%
\begin{wrapfigure}[12]{r}{3in}%
\includegraphics[width=3in]{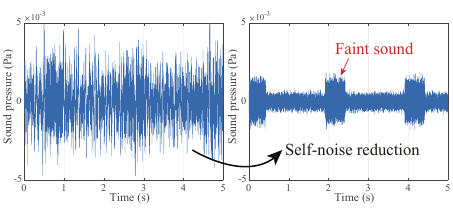}%
\end{wrapfigure}%
\begin{abstract}
The self-noise of capacitive sensors, primarily caused by thermal noise from the gate-bias resistor in the preamplifier, imposes a fundamental limit on measurement sensitivity. In electret condenser microphones (ECMs), this resistor simultaneously determines the noise low-pass cutoff frequency and the signal high-pass cutoff frequency through a single RC time constant, creating a trade-off between noise reduction and signal bandwidth. This paper proposes PDS-Amp (Photoelectric DC Servo Amplifier), a circuit technique that replaces the gate-bias resistor with a photoelectric element functioning as an ultra-high-impedance current source. A DC servo loop using lag-lead compensation feeds back the preamplifier output through an LED to control the photocurrent, thereby stabilizing the gate bias while decoupling the noise and signal cutoff frequencies. A custom photosensor based on the external photoelectric effect of a zinc photocathode was fabricated to achieve sub-picoampere dark current, overcoming the limitations of commercial semiconductor photodiodes. Combined with a cascode JFET preamplifier that minimizes input capacitance through bootstrap action, PDS-Amp achieved a self-noise of 11 dBA with a 12~pF dummy microphone. Despite using a small-diameter ECM capsule, this performance is comparable to that of large-diaphragm condenser microphones costing several thousand dollars. Recording experiments with an actual ECM capsule qualitatively confirmed a significant reduction in background noise. The proposed technique is applicable not only to microphones but broadly to capacitive sensors including accelerometers, pressure sensors, and pyroelectric sensors.
\end{abstract}
% 静電容量型センサのセルフノイズは、主にプリアンプ内のゲート・バイアス抵抗の熱雑音に起因し、計測感度の基本的な限界を規定する。エレクトレットコンデンサマイクロフォン（ECM）では、この抵抗が単一のRC時定数を通じてノイズの低域通過カットオフ周波数と信号の高域通過カットオフ周波数を同時に決定するため、ノイズ低減と信号帯域確保の間にトレードオフが存在する。本論文は、ゲート・バイアス抵抗を超高インピーダンスの電流源として機能する光電素子に置き換える回路技術PDS-Amp（Photoelectric DC Servo Amplifier）を提案する。ラグリード補償を用いたDCサーボループが、プリアンプ出力からLEDを経由して光電流を制御し、ゲートバイアスを安定化させつつ、ノイズと信号のカットオフ周波数を独立に設計可能とする。市販の半導体フォトダイオードの限界を克服するため、亜鉛光電陰極の外部光電効果を利用した自作光電センサを作製し、サブpAの暗電流を実現した。ブートストラップ動作により入力容量を最小化するカスコードJFETプリアンプとの組み合わせで、12 pFの疑似マイクにおいて11 dBAのセルフノイズを達成した。小口径のECMカプセルでありながら、この性能は数千ドルクラスのラージダイアフラムコンデンサマイクロフォンに匹敵する水準である。実ECMカプセルを用いた録音実験では、背景ノイズの大幅な低減を定性的に確認した。提案手法は、マイクロフォンに限らず、加速度センサ、圧力センサ、焦電センサを含む静電容量型センサに広く適用可能である。
\end{minipage}}}

\maketitle

%%%%%%%%%%%%%%%%%%%%%%%%%%%%%%%%%%%%%%%%%%%%%%%%%%%%%
%% Section 1: Introduction
%%%%%%%%%%%%%%%%%%%%%%%%%%%%%%%%%%%%%%%%%%%%%%%%%%%%%
\section{Introduction}

% 静電容量型センサとは、物理量を静電容量の変化で捉える方式のセンサである。この種のセンサは、変位、圧力、加速度といった多様な物理的変化を、静電容量の変化として変換し取り出す。その応用範囲は広く、加速度センサや圧力センサ、焦電センサなど、産業機器からコンシューマ製品に至るまで、現代の様々な技術分野で基盤的な役割を担っている。
\IEEEPARstart{C}{apacitive} sensors detect physical quantities by measuring changes in capacitance.
These sensors transduce a wide variety of physical changes---displacement, pressure, acceleration, and others---into corresponding variations in capacitance.
Their applications span numerous technical fields, from industrial equipment to consumer products, including accelerometers, pressure sensors, and pyroelectric sensors~\cite{Cheng-2023-RecentAdvancesCapacitiveSensors, Eargle-2004-MicrophoneBook}.

% これらの静電容量型センサの中でも、本研究では、一例としてコンデンサマイクロフォンにフォーカスする。コンデンサマイクロフォンは、代表的な静電容量型センサであり、音響信号を高感度に検出できる。その基本構造は、音波によって振動するダイヤフラムと固定電極から成り、両者間の距離の変化が静電容量の変化として生じる。特に、外部から固定電極に電圧付加が不要な、エレクトレットコンデンサマイクロフォン（ECM）は、その簡便さから広く普及している。
Among the various types of capacitive sensors, this study focuses on the condenser microphone as an illustrative example.
The condenser microphone is a representative capacitive sensor capable of detecting acoustic signals with high sensitivity.
Its basic structure consists of a diaphragm that vibrates in response to sound waves and a fixed back-electrode; variations in the gap between them produce changes in capacitance.
In particular, the electret condenser microphone (ECM)~\cite{Sessler-1963-ElectrostaticMicrophonesElectretFoil}, which requires no externally applied polarization voltage on the back-electrode, is widely adopted owing to its simplicity.

\begin{figure}[t]
\centering
\includegraphics[width=86mm]{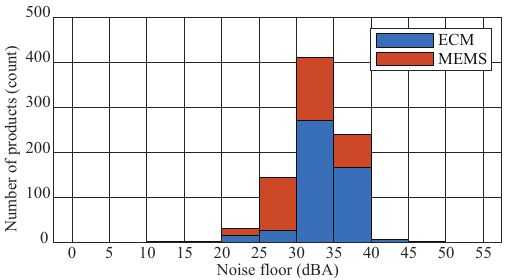}
\caption{Self-noise distribution of commercially available microphones. Histograms of the noise floor for ECMs (500~products) and MEMS microphones (347~products). The majority of products fall in the 30~dBA range, and products below 20~dBA are exceedingly rare.}
\label{fig_NFdBA}
\end{figure}

% 小型のコンデンサマイクロフォンは、大変便利な存在であるが、小型ゆえ構造上の性能限界が存在する。
Although miniature condenser microphones are highly convenient, their small size imposes an inherent performance limit.
Fig.~\ref{fig_NFdBA} presents a survey of the equivalent input noise level (self-noise) for a large number of commercially available ECM and MEMS microphones.
As the figure indicates, products with self-noise below 20~dBA are exceedingly rare, revealing a technical barrier for miniature microphones.
In the MEMS microphone domain, various approaches to improving the signal-to-noise ratio have been investigated through advances in sensor structure and readout circuit design~\cite{Zawawi-2020-ReviewMEMSCapacitiveMicrophones}.
Nevertheless, overcoming the 20~dBA barrier has traditionally required large diaphragms or high-voltage circuits, both of which conflict with miniaturization.

%%%
% NFdBA Figure

% マイクロフォンのセルフノイズは、計測対象の環境ノイズよりも十分に小さくなければならない。
A microphone's self-noise must be substantially lower than the ambient noise level of the target measurement environment.
If the self-noise is comparable to the ambient noise, the recorded noise floor increases by approximately 3~dB owing to the power addition of two uncorrelated noise sources.
For example, a microphone with 20~dBA self-noise cannot faithfully capture a 20~dBA acoustic environment because the sensor itself contributes a comparable amount of noise energy.
Therefore, microphones always require a self-noise level sufficiently below the ambient noise of the target measurement environment.

% この性能限界の要因を理解するには、マイクロフォンにおけるノイズを区別して考える必要がある。
Understanding the origin of this performance limit requires distinguishing between two types of microphone noise.
Microphone noise can be categorized into ``external noise,'' which is coupled from transmission lines and power supplies, and ``self-noise,'' which is physically generated by the sensor itself.
Prior research on microphone noise has primarily focused on suppressing external noise through techniques such as differential transmission; for example, the authors have previously reported a circuit technique that fundamentally suppresses external noise~\cite{Obo-2025-HighPSRRLowOutputb}.
Thus, external noise has been the main target in the existing literature.

% センサの計測限界を決めているのは、セルフノイズである。
The measurement limit of a sensor is ultimately determined by its self-noise.
Therefore, achieving challenging measurements demands a strong emphasis on reducing self-noise.
The dominant source of self-noise is the thermal noise generated by the resistive element in the bias circuit of the built-in preamplifier~\cite{Cordell-2024-DesigningAudioCircuitsSystems,Motchenbacher-1993-LowNoiseElectronicSystemDesign,Zuckerwar-2003-BackgroundNoisePiezoresistiveElectret}, as will be detailed in the following sections.
Conventionally, to compensate for the signal-to-noise ratio degradation caused by this thermal noise, approaches such as applying a high polarization voltage of approximately 200~V to the diaphragm or employing a large-area diaphragm have been adopted.
However, these approaches to enhancing signal sensitivity inevitably lead to larger sensor assemblies and surrounding circuits.

% そこで本稿で提案するセルフノイズ低減手法は、センサ内部のバイアス回路を工夫する、回路技術「PDS-Amp」である。
This paper proposes a self-noise reduction technique called PDS-Amp (Photoelectric DC Servo Amplifier), which modifies the internal bias circuit of the sensor.
The key idea is to replace the gate-bias resistor---the primary source of thermal noise---with a photoelectric element such as a photodiode or photoemissive cell.
This photoelectric element functions as a current source with extremely high impedance, and a DC servo loop actively stabilizes the bias voltage.
Through this configuration, PDS-Amp overcomes the constraints of conventional techniques and reduces the self-noise of capacitive sensors.

% 本論文の残りの部分は、以下の構成となっている。
The remainder of this paper is organized as follows.
Section~II presents the theoretical background of self-noise and details the operating principle of the proposed PDS-Amp.
Section~III describes the specific circuit design and implementation.
Section~IV demonstrates the effectiveness of PDS-Amp through experimental results obtained with a prototype microphone.
Finally, Section~V summarizes the contributions and discusses future directions.

%%%%%%%%%%%%%%%%%%%%%%%%%%%%%%%%%%%%%%%%%%%%%%%%%%%%%
%% Section 2: Theoretical Background and Proposed Method
%%%%%%%%%%%%%%%%%%%%%%%%%%%%%%%%%%%%%%%%%%%%%%%%%%%%%
\section{Theoretical Background and Proposed Method}
% 本セクションでは、まずコンデンサマイクロフォンのセルフノイズに関する理論的背景を述べ、次にセルフノイズを解決するためのPDS-Amp法の原理について説明する。
This section first describes the theoretical background of self-noise in condenser microphones and then explains the operating principle of the PDS-Amp method for reducing self-noise.

\subsection{Principle of Condenser Microphones}

\begin{figure*}[!t]
\centering
\includegraphics[width=177mm]{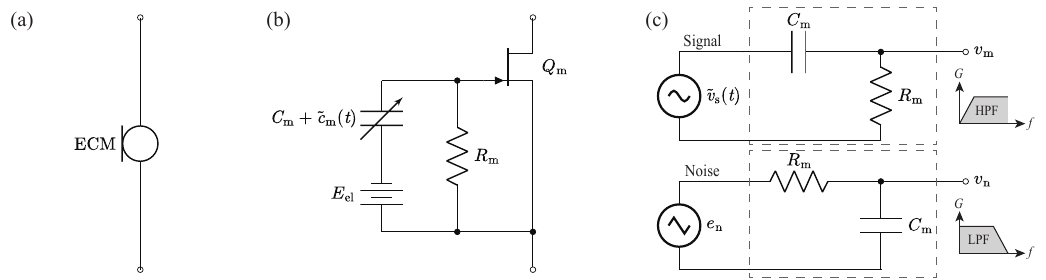}
\caption{Equivalent circuit of an ECM. (a)~Circuit symbol. (b)~Equivalent circuit (electret field $E_{\mathrm{el}}$, capacitance $C_{\mathrm{m}}+\tilde{c}_{\mathrm{m}}(t)$, gate-bias resistor $R_{\mathrm{m}}$, JFET $Q_{\mathrm{m}}$). (c)~Small-signal equivalent circuit (upper: signal equivalent circuit; lower: noise equivalent circuit).}
\label{fig_ECM_equiv}
\end{figure*}

% ECMは、エレクトレット材料により外部電源なしで動作する静電容量型センサである。
An ECM is a capacitive sensor that operates without an external power supply by virtue of its electret material~\cite{Sessler-1963-ElectrostaticMicrophonesElectretFoil, Eargle-2004-MicrophoneBook}.
Fig.~\ref{fig_ECM_equiv}(a) shows the circuit symbol of the ECM, and Fig.~\ref{fig_ECM_equiv}(b) shows its equivalent circuit.
The electret material retains a semi-permanent electric field $E_{\mathrm{el}}$ through a poling process, and when combined with a capacitor structure, it stores a charge $Q_0 = C_{\mathrm{m}} E_{\mathrm{el}}$ at rest.
Here, $C_{\mathrm{m}}$ is the quiescent capacitance of the capacitor formed by the diaphragm and the back-electrode.
This ability to operate without any externally supplied voltage or charge is a defining characteristic of the ECM.

%%%
% ECM_equiv Figure

% ECMの信号変換は、ダイヤフラム振動に伴う静電容量の変動に基づく。
The signal transduction of an ECM is based on the capacitance variation caused by diaphragm vibration.
As shown in Fig.~\ref{fig_ECM_equiv}(b), the quiescent capacitance is denoted by $C_{\mathrm{m}}$, and the small capacitance variation induced by diaphragm vibration due to sound waves is defined as $\tilde{c}_{\mathrm{m}}(t)$.
Hereafter, the instantaneous capacitance is expressed as $C_{\mathrm{m}}+\tilde{c}_{\mathrm{m}}(t)$.
In this paper, the small-signal condition $|\tilde{c}_{\mathrm{m}}(t)| \ll C_{\mathrm{m}}$ is assumed.

% ECMの静止点は、ゲート・バイアス抵抗$R_{\mathrm{m}}$によって電荷を定常状態に保つことで定義される。
The quiescent operating point of the ECM is established by the gate-bias resistor $R_{\mathrm{m}}$, which maintains the charge in a steady state.
The element $R_{\mathrm{m}}$ in Fig.~\ref{fig_ECM_equiv}(b) represents a gate-bias resistor on the order of G$\Omega$ housed inside the capsule.
Because $R_{\mathrm{m}}$ provides a DC path, the charge stored on the capacitor is held in a stable quiescent state without divergence of the potential.
In other words, $R_{\mathrm{m}}$ serves to keep the JFET gate voltage at a well-defined DC operating point.

% 可聴帯域の信号に対しては、$R_{\mathrm{m}}$が十分大きいため電荷一定近似が成立する。
For signals in the audible frequency range, $R_{\mathrm{m}}$ is sufficiently large that a constant-charge approximation holds.
Because $R_{\mathrm{m}}$ is on the order of G$\Omega$ and $C_{\mathrm{m}}$ is on the order of tens of picofarads, the $R_{\mathrm{m}}C_{\mathrm{m}}$ time constant reaches tens of milliseconds to several seconds.
Since the period of signals in the audible band (20~Hz--20~kHz) is shorter than this time constant, charge flow through $R_{\mathrm{m}}$ is negligible at signal frequencies.
Therefore, the capacitor charge can be approximated as $Q(t) \approx Q_0$, and the voltage across the diaphragm is given by $Q_0/(C_{\mathrm{m}}+\tilde{c}_{\mathrm{m}}(t))$.

% 電荷一定近似と小信号条件の下で、容量変動は一次の電圧変動に変換される。
Under the constant-charge approximation and the small-signal condition, the capacitance variation is converted into a first-order voltage variation.
Factoring $C_{\mathrm{m}}$ out of the denominator in the expression for the diaphragm voltage yields
\begin{equation}
\frac{Q_0}{C_{\mathrm{m}}+\tilde{c}_{\mathrm{m}}(t)}
=
\frac{Q_0}{C_{\mathrm{m}}}\cdot
\frac{1}{1+\tilde{c}_{\mathrm{m}}(t)/C_{\mathrm{m}}}.
\end{equation}
Applying the approximation $1/(1+x) \approx 1-x$ for $|\tilde{c}_{\mathrm{m}}(t)| \ll C_{\mathrm{m}}$ gives
\begin{equation}
\frac{Q_0}{C_{\mathrm{m}}+\tilde{c}_{\mathrm{m}}(t)}
\approx
\frac{Q_0}{C_{\mathrm{m}}}
-\frac{Q_0}{C_{\mathrm{m}}^2}\tilde{c}_{\mathrm{m}}(t).
\label{eq_dc_ac}
\end{equation}
The first term on the right-hand side is the DC component, and the second term is the AC component proportional to the capacitance variation $\tilde{c}_{\mathrm{m}}(t)$.

% 以上の導出から、ECMの小信号等価回路は等価電圧源$\tilde{v}_{\mathrm{s}}(t)$で表現される。
From the above derivation, the small-signal equivalent circuit of the ECM is represented by an equivalent voltage source $\tilde{v}_{\mathrm{s}}(t)$.
In the small-signal equivalent circuit, the DC source is treated as zero for AC analysis, so the DC component $Q_0/C_{\mathrm{m}}$ is excluded.
Consequently, only the AC component---the second term in (\ref{eq_dc_ac})---appears as the signal.
Fig.~\ref{fig_ECM_equiv}(c) depicts this AC component as an equivalent voltage source $\tilde{v}_{\mathrm{s}}(t)$, forming the small-signal equivalent circuit of the ECM.
The equivalent voltage source is expressed as
\begin{equation}
\tilde{v}_{\mathrm{s}}(t)\approx -\frac{Q_0}{C_{\mathrm{m}}^2}\tilde{c}_{\mathrm{m}}(t).
\label{eq_vs}
\end{equation}
Equation (\ref{eq_vs}) shows that the capacitance variation is linearly converted into a voltage variation, and it serves as the starting point for the subsequent noise analysis.

\subsection{Noise Spectrum Shaping by the Gate Resistor}

% ゲート・バイアス抵抗$R_{\mathrm{m}}$が発生する熱雑音は、ECMのセルフノイズの支配的な要因である。
The thermal noise generated by the gate-bias resistor $R_{\mathrm{m}}$ is the dominant source of self-noise in an ECM~\cite{Donk-1991-GeneralConsiderationsNoiseMicrophone}.
The lower part of Fig.~\ref{fig_ECM_equiv}(c) shows the noise equivalent circuit including the thermal noise of $R_{\mathrm{m}}$.
The resistor $R_{\mathrm{m}}$ possesses a white thermal noise voltage source $e_{\mathrm{n}}$ whose voltage spectral density is $\sqrt{4k_{\mathrm{B}}TR_{\mathrm{m}}}$~[V/$\sqrt{\mathrm{Hz}}$], where $k_{\mathrm{B}}$ is the Boltzmann constant and $T$ is the absolute temperature.
As seen from the gate, this thermal noise appears at the JFET input through the RC network formed by $R_{\mathrm{m}}$ and $C_{\mathrm{m}}$.
The following discussion describes how this RC network shapes the noise spectrum.

% $R_{\mathrm{m}}$と$C_{\mathrm{m}}$の並列RC回路は、熱雑音スペクトルに1次ローパス特性を与える。
The parallel RC network of $R_{\mathrm{m}}$ and $C_{\mathrm{m}}$ imposes a first-order low-pass characteristic on the thermal noise spectrum.
The low-pass filter transfer function of the RC network, in which $C_{\mathrm{m}}$ is connected in parallel with the thermal noise voltage source $e_{\mathrm{n}}$ of $R_{\mathrm{m}}$, is
\begin{equation}
H(f) = \frac{1}{1 + j2\pi f R_{\mathrm{m}} C_{\mathrm{m}}}.
\label{eq_Hf}
\end{equation}
Multiplying the voltage spectral density of the thermal noise source, $\sqrt{4k_{\mathrm{B}}TR_{\mathrm{m}}}$~[V/$\sqrt{\mathrm{Hz}}$], by $H(f)$ yields the voltage noise spectral density at the gate terminal, $v_{\mathrm{n}}(f)$:
\begin{equation}
v_{\mathrm{n}}(f) = \frac{\sqrt{4k_{\mathrm{B}}TR_{\mathrm{m}}}}{1 + j2\pi f R_{\mathrm{m}} C_{\mathrm{m}}}.
\label{eq_noise_nsd}
\end{equation}
The magnitude $|v_{\mathrm{n}}(f)|$ in (\ref{eq_noise_nsd}) exhibits a first-order low-pass filter characteristic with a cutoff frequency $f_{\mathrm{c}} = 1/(2\pi R_{\mathrm{m}} C_{\mathrm{m}})$.
Below $f_{\mathrm{c}}$, the spectral density is flat at $\sqrt{4k_{\mathrm{B}}TR_{\mathrm{m}}}$~[V/$\sqrt{\mathrm{Hz}}$], whereas above $f_{\mathrm{c}}$ it rolls off at $-20$~dB/dec~\cite{Cordell-2024-DesigningAudioCircuitsSystems}.

\begin{figure}[t]
\centering
\includegraphics[width=86mm]{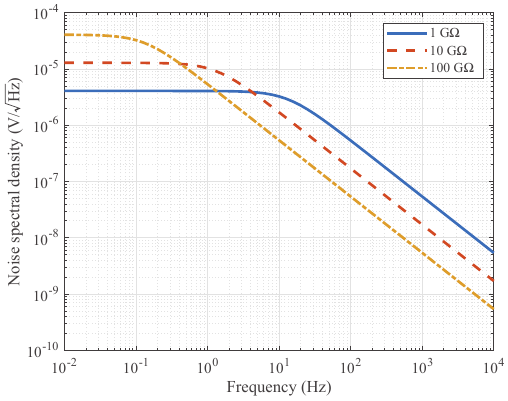}
\caption{Calculated noise spectral density as a function of the gate-bias resistance. With $C_{\mathrm{m}}=12$~pF, three conditions are shown: $R_{\mathrm{m}}=1$~G$\Omega$ (solid), $10$~G$\Omega$ (dashed), and $100$~G$\Omega$ (dash-dotted). A larger $R_{\mathrm{m}}$ concentrates the noise energy into a lower frequency band, reducing the spectral density in the audible range.}
\label{fig_ThermalNoise}
\end{figure}

% 抵抗値$R_{\mathrm{m}}$を大きくすると、ノイズスペクトル密度が可聴帯域で低下する理由は、ノイズの総電力が一定のままスペクトル形状が変化するためである。
The reason why increasing $R_{\mathrm{m}}$ lowers the noise spectral density in the audible band is that the total noise power remains constant while the spectral shape changes.
Integrating $|v_{\mathrm{n}}(f)|^2$ over all frequencies gives the total noise power:
\begin{equation}
\int_0^{\infty} |v_{\mathrm{n}}(f)|^2\,\mathrm{d}f = \frac{k_{\mathrm{B}}T}{C_{\mathrm{m}}},
\label{eq_total_noise}
\end{equation}
which is independent of $R_{\mathrm{m}}$.
This means that although the total noise energy is invariant regardless of $R_{\mathrm{m}}$, a larger $R_{\mathrm{m}}$ lowers the cutoff frequency $f_{\mathrm{c}}$, thereby concentrating that energy into a lower frequency band.
Fig.~\ref{fig_ThermalNoise} shows the noise spectral density calculated for $C_{\mathrm{m}}=12$~pF with $R_{\mathrm{m}}$ set to 1~G$\Omega$, 10~G$\Omega$, and 100~G$\Omega$.
It can be confirmed that a larger $R_{\mathrm{m}}$ results in a lower spectral density in the audible band (above 20~Hz).

%%%
% ThermalNoise Figure

% しかし、ECMの$R_{\mathrm{m}}C_{\mathrm{m}}$回路では、ノイズの低域通過カットオフと信号の高域通過カットオフが同一の時定数で決まるトレードオフが存在する。
However, in the $R_{\mathrm{m}}C_{\mathrm{m}}$ circuit of an ECM, there exists a trade-off in which the low-pass cutoff for noise and the high-pass cutoff for the signal are determined by the same time constant (Fig.~\ref{fig_ECM_equiv}(c)).
As described above, the $R_{\mathrm{m}}C_{\mathrm{m}}$ time constant determines the cutoff frequency $f_{\mathrm{c}}$ of the low-pass filter for thermal noise.
Simultaneously, this $f_{\mathrm{c}}$ also functions as the lower cutoff frequency of the high-pass characteristic for the signal.
% This is because signal components at frequencies below $f_{\mathrm{c}}$ are attenuated as the charge relaxes through $R_{\mathrm{m}}$ and does not appear at the output.
In other words, the two cutoff frequencies---the low-pass filter that pushes noise into the low-frequency region and the high-pass filter that passes the signal---are both constrained to the single value $f_{\mathrm{c}} = 1/(2\pi R_{\mathrm{m}} C_{\mathrm{m}})$.

% このトレードオフにより、$R_{\mathrm{m}}$の増大のみによるノイズ低減には実用上の限界がある。
Because of this trade-off, noise reduction solely by increasing $R_{\mathrm{m}}$ has practical limitations.
Increasing $R_{\mathrm{m}}$ lowers $f_{\mathrm{c}}$, shifting both the noise low-pass cutoff and the signal high-pass cutoff simultaneously toward lower frequencies.
Passing audible-band signals without degradation requires $f_{\mathrm{c}} < 20$~Hz, which for $C_{\mathrm{m}}=12$~pF corresponds to $R_{\mathrm{m}} > 0.6$~G$\Omega$.
For example, at $R_{\mathrm{m}} = 1$~G$\Omega$, $f_{\mathrm{c}} \approx 13$~Hz and the impact on the signal band is small, but the noise reduction in the audible band is also limited.
Increasing $R_{\mathrm{m}}$ to 10~G$\Omega$ yields $f_{\mathrm{c}} \approx 1.3$~Hz and further reduces the noise spectral density; however, the $R_{\mathrm{m}}C_{\mathrm{m}}$ time constant then reaches several seconds, requiring a long settling time before the gate voltage converges to a stable quiescent state.
Furthermore, at high resistance values, impedance variations due to temperature and humidity become significant, causing gate-bias drift or divergence that compromises practical stability.
PDS-Amp is a method that fundamentally resolves this trade-off by enabling independent design of the noise low-pass cutoff frequency and the signal high-pass cutoff frequency.

\subsection{Concept of the Photoelectric DC Servo in PDS-Amp}

% 前節で述べたトレードオフの根本原因は、$R_{\mathrm{m}}$が、ノイズ源とDCバイアス供給と信号のハイパスフィルタの役割を同時に担っている点にある。
The fundamental cause of the trade-off described in the preceding subsection is that $R_{\mathrm{m}}$ simultaneously serves as a noise source, a DC bias supply path, and a signal high-pass filter.
Because $R_{\mathrm{m}}$ is a physical resistor, it inevitably generates thermal noise.
At the same time, $R_{\mathrm{m}}$ is the bias path that establishes the DC potential of the JFET gate, and the $R_{\mathrm{m}}C_{\mathrm{m}}$ time constant also determines the low-frequency cutoff of the signal.
The fact that these multiple roles are consolidated in a single element makes it difficult to simultaneously achieve noise reduction and signal-band preservation.

% このトレードオフを解消するために、PDS-Ampでは$R_{\mathrm{m}}$を電流源に置き換える。
To resolve this trade-off, PDS-Amp replaces $R_{\mathrm{m}}$ with a current source.
Several approaches to noise reduction in preamplifier bias circuits have been reported in the literature.
For example, a charge amplifier that replaces the resistor with a capacitor is known, but this configuration suffers from DC bias divergence~\cite{Bertuccio-1993-NovelChargeSensitivePreamplifier}.
In MEMS microphone readout circuits, a technique has been reported in which an internal feedback path is added to the source-follower input buffer to suppress the noise contribution of the bias current-source circuit by loop gain~\cite{Li-2022-AnalogReadoutCircuitNoiseReductiona}.
In contrast, the present work reconfigures the bias path itself by replacing the physical resistor---the primary source of thermal noise---with a current source.
An ideal current source has infinite impedance, enabling the noise low-pass cutoff frequency as seen from the gate to be set extremely low.
Meanwhile, because a current source functions differently from a resistor as a DC bias path, the signal high-pass cutoff frequency can be set independently by a separate mechanism.
In other words, replacing the resistor with a current source fundamentally decouples the noise characteristic from the signal characteristic.

% PDS-Ampでは、この電流源として光電素子を採用する。
In PDS-Amp, a photoelectric element is employed as this current source.
A photoelectric element generates current only upon illumination and draws virtually no current in the dark.
It can therefore be regarded as a light-controllable current source.
By adjusting the illumination intensity, picoampere-order currents can be precisely controlled, providing the current necessary to adjust the gate bias.
A prior study using optical feedback has been reported for a current amplifier~\cite{Higa-2006-NoiseAnalysisVeryLowlevel}; however, the underlying concept differs from that of the capacitive-sensor preamplifier in the present work.

% また、PDS-Ampでは、DCサーボ回路を併用する。
In addition, PDS-Amp incorporates a DC servo circuit.
As described in the preceding subsection, $R_{\mathrm{m}}$ served the role of stabilizing the JFET gate voltage at DC.
Unlike a resistor, the photoelectric element used as a current source has no inherent bias-stabilizing function; any slight variation in the charge accumulated on the gate directly manifests as gate-voltage drift.
PDS-Amp addresses this drift problem by introducing a negative feedback loop via light (a DC servo).

\begin{figure}[t]
\centering
\includegraphics[width=86mm]{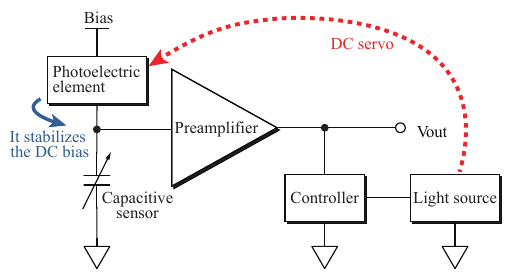}
\caption{Conceptual block diagram of PDS-Amp. A photoelectric element and the capacitive sensor are both connected to the preamplifier gate, and a DC servo loop feeds back from the preamplifier output through a lag-lead controller and an LED light source to the photoelectric element.}
\label{fig_CircuitBlock}
\end{figure}

% このDCサーボループは、プリアンプ出力から光電素子への照射光量を制御し、ゲートバイアスを安定化させる。
This DC servo loop controls the illumination intensity directed at the photoelectric element from the preamplifier output, thereby stabilizing the gate bias.
Fig.~\ref{fig_CircuitBlock} shows the block diagram of PDS-Amp.
The preamplifier output voltage is fed to a controller (lag-lead compensator), whose output controls the LED drive current.
The LED light illuminates the photoelectric element, which generates a small current based on the photoelectric effect; this current adjusts the gate voltage.
If the gate voltage rises above the target value, the preamplifier output changes, the LED intensity is adjusted, the photoelectric element current changes accordingly, and the gate voltage is pulled back toward the target.
Through this closed-loop operation, the DC bias voltage at the gate is stabilized at the target value.

% DCサーボの制御器には、ラグリード補償を採用することで、ループの発振を防止する。
Lag-lead compensation is adopted for the DC servo~\cite{Cordell-2024-DesigningAudioCircuitsSystems} controller to prevent loop oscillation.
The capacitive sensor itself is a capacitor, and the charge-storage action causes the phase to approach $-90^\circ$ at low frequencies, giving it an integrating characteristic.
If a pure integrator were used as the DC servo controller, the integrating characteristic of the sensor ($-90^\circ$) and that of the controller ($-90^\circ$) would add up, resulting in insufficient phase margin and oscillation.
The lag-lead compensator maintains high loop gain at low frequencies like an integrator, while boosting the phase near the gain-crossover frequency to ensure adequate phase margin.
In this work, the closed-loop high-pass filter (HPF) cutoff frequency is set in the range of 10--20~Hz, which has virtually no effect on signal transmission in the audible band.

%%%
% CircuitBlock Figure

% なお、高インピーダンスのバイアス素子としてはトランジスタ定電流回路も候補となるが、ノイズ電流の制約から適さない。
Although a transistor-based constant-current circuit is also a candidate for a high-impedance bias element, it is unsuitable because of its noise-current limitation.
In a transistor-based constant-current circuit, shot noise and $1/f$ noise from carriers in the channel typically produce a noise-current density on the order of several pA/$\sqrt{\mathrm{Hz}}$.
When this noise current $i_{\mathrm{n}}$ flows into $C_{\mathrm{m}}$, it produces a voltage noise of $i_{\mathrm{n}}/(2\pi f C_{\mathrm{m}})$ at the gate terminal.
For example, with $i_{\mathrm{n}} = 5$~pA/$\sqrt{\mathrm{Hz}}$, $C_{\mathrm{m}}=12$~pF, and $f = 1$~kHz, the gate-terminal voltage noise is approximately 66~$\mu$V/$\sqrt{\mathrm{Hz}}$, which far exceeds the input-referred noise of the JFET (a few nV/$\sqrt{\mathrm{Hz}}$).
Therefore, the noise current of a transistor constant-current circuit is far too large relative to the capacitance of the ECM.

% 物理抵抗を電流源に置き換えた場合、熱雑音は消失するが、代わりに電流源素子由来のショットノイズが発生する。
When a physical resistor is replaced by a current source, the thermal noise vanishes but is replaced by shot noise originating from the current-source element.
The shot-noise current density of an element carrying a current $I$ is~\cite{Motchenbacher-1993-LowNoiseElectronicSystemDesign}
\begin{equation}
\sqrt{2qI} \quad [\mathrm{A}/\sqrt{\mathrm{Hz}}],
\label{eq_2qI}
\end{equation}
where $q$ is the elementary charge.
For example, assuming a bias current of $I = 1$~pA, the shot-noise current density is $\sqrt{2 \times 1.6 \times 10^{-19} \times 10^{-12}} \approx 0.57$~fA/$\sqrt{\mathrm{Hz}}$.
In contrast, the thermal-noise current density of a 1~G$\Omega$ resistor is $\sqrt{4k_{\mathrm{B}}T/R} \approx 4.1$~fA/$\sqrt{\mathrm{Hz}}$, and even for 10~G$\Omega$ it is approximately 1.3~fA/$\sqrt{\mathrm{Hz}}$.
Thus, in the picoampere-order regime, the shot noise of the current source (0.57~fA/$\sqrt{\mathrm{Hz}}$) is only about one-seventh of the thermal noise of a 1~G$\Omega$ resistor and about half that of a 10~G$\Omega$ resistor.
Therefore, replacing a physical resistor with a current source shifts the dominant noise mechanism from thermal noise to shot noise, thereby improving the physical limit of the signal-to-noise ratio.

% なお、光電素子のショットノイズがさらに低減された場合、次にノイズの下限を決定するのはプリアンプの入力バイアス電流である。
If the shot noise of the photoelectric element is further reduced, the next factor that determines the noise floor is the input bias current of the preamplifier.
A sub-picoampere leakage current (gate current) flows through the JFET gate, and the DC servo maintains the bias point through the balance between this gate current and the photoelectric element current.
For example, the maximum gate current of the JFE2140 is 10~pA~\cite{TI-JFE2140}; however, it is suppressed to the sub-picoampere order by keeping the drain--source voltage low through the cascode connection.
This gate current becomes the dominant factor of the new noise floor.
In other words, the ultimate lower limit of self-noise in PDS-Amp is determined not by the noise of the photoelectric element but by the input bias current of the preamplifier.

% 以上の構成により、PDS-Ampはノイズのローパスカットオフ周波数と信号のハイパスカットオフ周波数を独立に設計可能とする。
Through the above configuration, PDS-Amp enables independent design of the noise low-pass cutoff frequency and the signal high-pass cutoff frequency.
The extremely high equivalent impedance of the photoelectric element substantially reduces thermal noise and sets the noise low-pass cutoff at an extremely low frequency.
Simultaneously, the DC servo loop independently sets the signal high-pass cutoff frequency to 10--20~Hz, transmitting audible-band signals without degradation.
As a result, the coupling between the two cutoff frequencies---which was inseparable in the conventional $R_{\mathrm{m}}C_{\mathrm{m}}$ circuit---is fundamentally severed.

%%%%%%%%%%%%%%%%%%%%%%%%%%%%%%%%%%%%%%%%%%%%%%%%%%%%%
%% Section 3: Circuit Design and Implementation
%%%%%%%%%%%%%%%%%%%%%%%%%%%%%%%%%%%%%%%%%%%%%%%%%%%%%
\section{Circuit Design and Implementation of the PDS-Amp}
% 本セクションでは、PDS-Ampを具現化する具体的な回路設計と実装について述べる。
This section describes the specific circuit design and implementation that realize PDS-Amp.

\subsection{Overall Circuit Architecture}

% PDS-Ampの回路は、プリアンプ部とDCサーボループ部の2つのブロックで構成される。
The PDS-Amp circuit consists of two blocks: a preamplifier section and a DC servo loop section.
The preamplifier section amplifies the weak voltage signal from the capacitive sensor with low noise.
The DC servo loop section is composed of the photoelectric element, the lag-lead compensator, and the LED light source described in the preceding section, and it stabilizes the DC bias at the gate.
The entire circuit is powered by two series-connected 006P (9~V) batteries, providing a $\pm 9$~V supply.
Battery operation eliminates noise coupled from mains power supplies.

\subsection{Low-Noise Preamplifier Design}

% 静電容量型センサの出力電圧は、プリアンプの入力容量$C_{\mathrm{in}}$によって減衰する。
The output voltage of a capacitive sensor is attenuated by the input capacitance $C_{\mathrm{in}}$ of the preamplifier.
The equivalent voltage source $\tilde{v}_{\mathrm{s}}(t)$ of the ECM is connected to the preamplifier gate through the internal capacitance $C_{\mathrm{m}}$.
If the preamplifier has an input capacitance $C_{\mathrm{in}}$, then $C_{\mathrm{m}}$ and $C_{\mathrm{in}}$ form a capacitive voltage divider, and the signal voltage appearing at the gate is attenuated to
\begin{equation}
v_{\mathrm{g}} = \tilde{v}_{\mathrm{s}}(t) \cdot \frac{C_{\mathrm{m}}}{C_{\mathrm{m}} + C_{\mathrm{in}}}.
\label{eq_cap_div}
\end{equation}
For an ECM with $C_{\mathrm{m}} = 12$~pF, signal attenuation becomes non-negligible when $C_{\mathrm{in}}$ exceeds a few picofarads.
Therefore, minimizing the effective input capacitance of the preamplifier is essential for recording the output of the capacitive sensor without attenuation~\cite{Yoo-2021-ReadoutCircuitsCapacitiveSensors}.

% プリアンプの入力容量の主な原因はミラー効果であり、カスコード接続によって抑制できる。
The primary cause of preamplifier input capacitance is the Miller effect, which can be suppressed by a cascode connection~\cite{Abidi-1988-OperationCascodeGainStages}.
In a single-stage amplifier, the gate--drain parasitic capacitance $C_{\mathrm{gd}}$ is magnified by the voltage gain $A_{\mathrm{v}}$ through the Miller effect, increasing the effective input capacitance to $C_{\mathrm{gs}} + (1+A_{\mathrm{v}})C_{\mathrm{gd}}$.
In a cascode connection, the source of the output-stage Q2 is connected to the drain of the input-stage Q1, and a fixed voltage is applied to the gate of Q2.
This configuration clamps the drain voltage of Q1 to the gate voltage of Q2 minus its gate--source voltage, keeping it nearly constant.
Because the drain voltage swing of Q1 is suppressed, the current flowing through $C_{\mathrm{gd}}$ is reduced and the Miller effect is mitigated.

\begin{figure}[t]
\centering
\includegraphics[width=86mm]{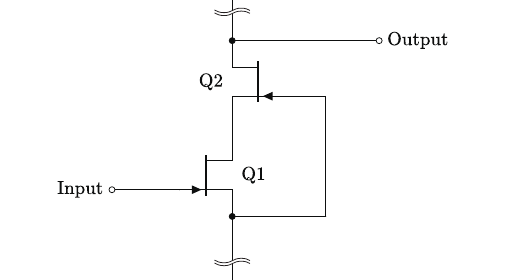}
\caption{Cascode circuit configuration using two JFE2140 devices. The source of the output-stage Q2 is connected to the drain of the input-stage Q1, clamping the drain voltage of Q1 to suppress the Miller effect. Simultaneously, bootstrap action reduces the effective input capacitance.}
\label{fig_Cascode}
\end{figure}

% さらに、カスコード接続はQ1に対するブートストラップとしても機能し、見かけ上の入力容量を低減する。
Furthermore, when connected as shown in Fig.~\ref{fig_Cascode}, the cascode connection also functions as a bootstrap for Q1, reducing the apparent input capacitance.
Because Q2 causes its source to follow the source voltage of Q1, the variation in the gate--drain potential difference of Q1 is minimized.
This behavior is equivalent to the bootstrap principle, in which the drain voltage tracks the gate voltage variation, suppressing charge transfer through $C_{\mathrm{gd}}$.
As a result, the effective input capacitance is reduced and the signal attenuation due to the capacitive divider in (\ref{eq_cap_div}) is minimized.

%%%
% Cascode Figure

% 本プリアンプでは、低ノイズJFET JFE2140を入力段と出力段の両方に使用した。
In the present preamplifier, the low-noise JFET JFE2140~\cite{TI-JFE2140, Featherstone-2023-JFE2140UltraLowNoisePreAmplifier} is used for both the input and output stages.
The JFE2140 has an input-referred noise of a few nV/$\sqrt{\mathrm{Hz}}$, making it well suited for amplifying the weak signals of capacitive sensors~\cite{Vallicelli-2020-3nVHzInputreferrednoiseAnalog}.
The input-stage Q1 operates as a source follower to receive the signal, while Q2 serves as the cascode stage to fix the drain voltage.
This configuration is not intended to extend the bandwidth but rather to prevent the signal attenuation due to capacitive voltage division shown in (\ref{eq_cap_div}), thereby faithfully recording the output of the capacitive sensor.

% ここで、JFE2140のパラメータを用いてカスコード接続の入力容量低減効果を定量的に示す。
The input-capacitance reduction achieved by the cascode connection is now quantified using the parameters of the JFE2140.
According to the JFE2140 datasheet~\cite{TI-JFE2140}, the input capacitance at $V_{\mathrm{DS}} = 5$~V is $C_{\mathrm{ISS}} = 13$~pF.
Since $C_{\mathrm{ISS}} = C_{\mathrm{gs}} + C_{\mathrm{gd}}$ and the feedback capacitance is $C_{\mathrm{gd}} \approx 1.2$~pF, the gate--source capacitance is estimated as $C_{\mathrm{gs}} \approx 11.8$~pF.
Assuming a single-stage amplifier with a voltage gain of $A_{\mathrm{v}} = 10$, the effective input capacitance including the Miller effect is $C_{\mathrm{gs}} + (1+A_{\mathrm{v}})C_{\mathrm{gd}} = 11.8 + 13.2 = 25.0$~pF.
In this case, the capacitive divider ratio for $C_{\mathrm{m}} = 12$~pF is $12/(12+25) \approx 0.32$ ($-9.8$~dB), meaning approximately two-thirds of the signal is lost.
With the cascode connection suppressing the Miller effect, the effective input capacitance is reduced to $C_{\mathrm{gs}} + C_{\mathrm{gd}} \approx 13.0$~pF, improving the divider ratio to $12/(12+13) = 0.48$ ($-6.4$~dB).
In a constant-current cascode connection without bootstrap (where the drain potential is fixed and the source operates under constant current), $V_{\mathrm{gs}}$ is held constant so that charging and discharging currents into $C_{\mathrm{gs}}$ are suppressed; however, because the drain does not follow the output signal, $C_{\mathrm{gd}}$ remains as input capacitance.
In this case, the effective input capacitance is $C_{\mathrm{gd}} \approx 1.2$~pF, and the divider ratio improves to $12/(12+1.2) \approx 0.91$ ($-0.8$~dB).
Also, when bootstrap action operates ideally, both $V_{\mathrm{gs}}$ and $V_{\mathrm{ds}}$ remain constant, so that $V_{\mathrm{gd}}$ is also held constant.
Consequently, charging and discharging currents into both $C_{\mathrm{gs}}$ and $C_{\mathrm{gd}}$ are suppressed, and they contribute negligibly to the input capacitance.
Under this condition, the sensor output can be extracted with virtually no attenuation.

\subsection{DC Servo Loop Implementation}

% 光電素子には、市販のフォトダイオードでは性能が不十分であった。
Commercial photodiodes proved insufficient for the photoelectric element.
In PDS-Amp, the DC servo maintains the bias point through the balance between the gate leakage current of the input JFET and the current from the photoelectric element.
The measured gate leakage current of the JFE2140 used in the present preamplifier was approximately 0.4~pA under the cascode connection.
Among the commercially available photodiodes with the lowest dark current, the Hamamatsu S5973-01~\cite{Hamamatsu-S5973-01} was considered; however, its datasheet specifies a typical dark current of 1~pA (at $V_{\mathrm{R}} = 10$~V).
Because the typical dark current of the S5973-01 (1~pA) exceeds the gate leakage current of the JFE2140 (0.4~pA), the photoelectric element current could not be balanced against the gate leakage current.
Even in the completely dark state with no illumination, the dark current from the photodiode exceeds the gate leakage current, making DC-servo-based bias adjustment infeasible.
Furthermore, the dark current exhibits unit-to-unit variation reaching up to 100~pA and has a strong temperature dependence; therefore, semiconductor photodiodes were judged unsuitable as the photoelectric element for PDS-Amp.

% PDS-Ampの光電素子に求められる要件は、低ノイズ電流と微小電流の制御性であり、光感度は重要ではない。
The requirements for the photoelectric element in PDS-Amp are low noise current and fine controllability of small currents; photosensitivity is not critical.
The current required for gate-bias adjustment by the DC servo is on the picoampere order, so high photosensitivity is not necessary.
What matters is that the leakage current in the dark state is extremely small and that the element provides stable current controllability with respect to the illumination level.
Therefore, an element that does not rely on a semiconductor PN junction is desirable, and a photoelectric element based on the external photoelectric effect is a strong candidate that meets these requirements.

\begin{figure}[t]
\centering
\includegraphics[width=86mm]{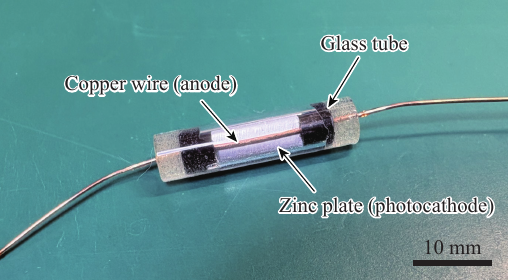}
\caption{Appearance and structure of the custom photosensor. A zinc plate (photocathode) and copper wire (anode) are housed in a 25~mm fuse-tube form factor and illuminated by a UV-C LED (275~nm). The interior is sealed and subjected to oxygen-removal treatment.}
\label{fig_PhotoSensor}
\end{figure}

% そこで、外部光電効果を利用する光電センサを自作した。
Accordingly, a custom photosensor utilizing the external photoelectric effect was fabricated.
Fig.~\ref{fig_PhotoSensor} shows the structure of the custom photosensor.
The custom photosensor has a structure in which electrons are emitted from a zinc plate via the external photoelectric effect upon ultraviolet irradiation.
A zinc plate serves as the photocathode and copper wire as the anode, both enclosed in a 25~mm glass tube (based on a fuse-tube form factor).
The work function of zinc is approximately 3.7--4.3~eV~\cite{Michaelson-1977-WorkFunctionElementsIts}, which is below the photon energy of the UV-C LED (wavelength 275~nm, photon energy approximately 4.5~eV); therefore, electrons are emitted by the photoelectric effect upon ultraviolet irradiation~\cite{Srinivasan-Rao-1991-PhotoemissionStudiesMetalsUsing}.
The glass tube is sealed with epoxy resin and has been subjected to oxygen-removal treatment.
Because electron emission based on the external photoelectric effect does not involve a semiconductor PN junction, there is no dark current, and the noise current is inherently extremely small.

%%%
% PhotoSensor Figure

% DCサーボループの補償器は、閉ループHPFカットオフが10--20~Hzとなるようにラグリード補償を設計した。
The DC servo loop compensator was designed with lag-lead compensation so that the closed-loop HPF cutoff falls in the range of 10--20~Hz.
The design guidelines were to ensure both loop stability (sufficient phase margin) and an HPF cutoff frequency that does not attenuate signals in the audible band.
The resulting closed-loop HPF cutoff frequency falls within the 10--20~Hz range, allowing acoustic signals above 20~Hz to be transmitted without attenuation.
The overall bandwidth of the preamplifier including the DC servo was measured to be approximately 2~MHz, which amply covers the audible band.

%%%%%%%%%%%%%%%%%%%%%%%%%%%%%%%%%%%%%%%%%%%%%%%%%%%%%
%% Section 4: Experimental Results
%%%%%%%%%%%%%%%%%%%%%%%%%%%%%%%%%%%%%%%%%%%%%%%%%%%%%

\section{Experimental Results}
% 本セクションでは、PDS-Ampの有効性を以下の3つの実験によって実証する。
This section demonstrates the effectiveness of PDS-Amp through the following three experiments.

\subsection{Measurement Setup}

% 本研究では、PDS-Ampの性能を評価するために3種類の実験を実施した。
Three types of experiments were conducted to evaluate the performance of PDS-Amp.
The outline of each experiment is as follows:
\begin{itemize}
\item \textbf{Test~1 (NSD comparison):} Using a dummy microphone, the noise spectral density of the conventional method (1~G$\Omega$ bias resistor) and PDS-Amp are compared inside a shielded enclosure.
\item \textbf{Test~2 (dBA evaluation):} The noise floor of the dummy microphone is quantified in dBA~SPL to evaluate the absolute self-noise level.
\item \textbf{Test~3 (recording experiment):} PDS-Amp is applied to an actual ECM capsule, and the qualitative effect of self-noise reduction is confirmed by recording a faint sound source in an anechoic chamber.
\end{itemize}

% Test~1およびTest~2で使用する疑似マイクは、MLCC（積層セラミックコンデンサ）で構成した。
The dummy microphone used in Test~1 and Test~2 was constructed from MLCCs (multilayer ceramic capacitors).
A dummy microphone is a substitute for an actual ECM capsule, in which an MLCC having the same capacitance (12~pF) serves as the sensor.
The MLCC used was composed of three 36~pF capacitors with C0G (NP0) characteristics connected in series to form 12~pF.
The reason for the series connection is that a single 12~pF MLCC had insufficient insulation resistance to maintain the required high impedance.
A shielded enclosure was used to block external electromagnetic noise, creating an environment in which only the noise of the circuit itself is evaluated.

% また、比較条件は、同一のプリアンプに対してバイアス供給部のみを変更する方法で統一した。
The comparison conditions were unified by changing only the bias supply section for the same preamplifier.
Under the PDS-Amp condition, the custom photosensor and the DC servo loop were used for bias supply.
Under the conventional condition, a 1~G$\Omega$ physical resistor was connected as the gate-bias resistor to the same preamplifier.
The preamplifier (JFE2140 cascode), the power supply ($\pm 9$~V battery), and the dummy microphone (MLCC 12~pF) were all common to both conditions.

% Test~1のNSD測定には、Cosmos ADCisoとREWを使用した。
The NSD measurements in Test~1 were performed using a Cosmos ADCiso and REW.
The Cosmos ADCiso is a USB-isolated ADC, and REW is an FFT analyzer and recorder software.
This combination directly outputs the vertical axis of the NSD in voltage spectral density [V/$\sqrt{\mathrm{Hz}}$].
Because no volume-based recording-gain adjustment is required as with a PCM recorder, the risk of measurement error is low.

% Test~2のdBA算出には、SONY PCM-D50とMATLABを使用した。
The dBA~\cite{IEC-61672-1-2013} evaluation in Test~2 was performed using a SONY PCM-D50 and MATLAB.
The procedure was as follows.
First, sensitivity calibration was performed using an ECM (C9767BB422LFP~\cite{CUI-C9767BB422LFP}, measured capacitance 12~pF) and an acoustic calibrator (B\&K Type 4231~\cite{BK-Type4231}, 94~dB SPL), and the reference level was recorded.
Next, the ECM was replaced with the 12~pF dummy microphone, and the noise floor was recorded.
Through this two-stage measurement, the noise floor of the dummy microphone was quantified in dBA SPL.

\subsection{Test~1: Noise Spectrum Density Comparison}

\begin{figure}[t]
\centering
\includegraphics[width=86mm]{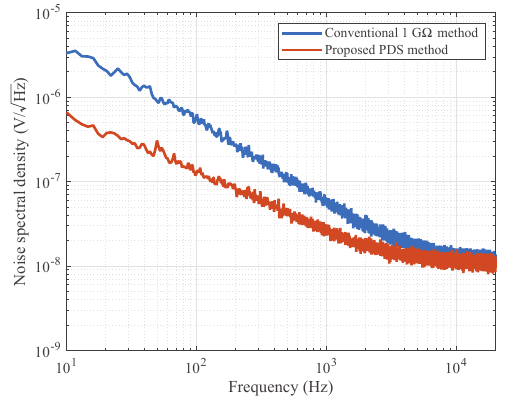}
\caption{Noise spectral density comparison between the conventional method (1~G$\Omega$ bias resistor) and PDS-Amp (custom photosensor). Measured with the dummy microphone (MLCC 12~pF) inside a shielded enclosure. PDS-Amp exhibits a lower noise spectral density over the entire band.}
\label{fig_CompPSD}
\end{figure}

% NSD比較において、PDS-Ampは可聴帯域全体で従来法を大幅に下回るノイズスペクトル密度を示した。
In the NSD comparison, PDS-Amp exhibited a substantially lower noise spectral density than the conventional method across the entire audible band.
Fig.~\ref{fig_CompPSD} shows the comparison of the noise spectral density between the conventional method (1~G$\Omega$) and PDS-Amp (custom photosensor).
The measurement bandwidth is 10~Hz--20~kHz, and PDS-Amp achieves a lower noise spectral density than the conventional method over the entire band.
The improvement is particularly pronounced in the low-frequency region (10--1000~Hz), confirming that noise in the band where thermal noise of the bias resistor was dominant has been effectively suppressed.

%%%
% CompPSD Figure

% PDS-AmpのNSD曲線では、低周波域から約1~kHz付近にかけて約$-10$~dB/decの傾斜が観察された。
The NSD curve of PDS-Amp exhibited a slope of approximately $-10$~dB/dec from the low-frequency region up to around 1~kHz.
In the conventional $R_{\mathrm{m}}C_{\mathrm{m}}$ circuit, the $-20$~dB/dec low-pass characteristic is dominant; however, under the PDS-Amp condition, a different slope of $-10$~dB/dec appeared.
This $-10$~dB/dec slope may be attributable to the $1/f$ noise component of the JFET, but there is also the possibility of a combined contribution from flicker noise of the photoelectric element and the frequency characteristic of the DC servo loop, and the origin has not been definitively identified at this time.
A similar slope characteristic in a cascode-connected JFET preamplifier has also been reported in a prior study~\cite{Jefferts-1987-LowNoiseCascodeAmplifier}.
In that report, a downward-sloping noise spectrum from the low-frequency region was observed in a low-noise preamplifier using cascode-connected 2SK117 devices, confirming a frequency dependence similar to that of the present work.
Elucidating the physical origin of the $-10$~dB/dec slope in the PDS-Amp NSD curve remains a subject for future investigation.

\subsection{Test~2: Self-Noise Evaluation in dBA}

% dBA評価において、PDS-Ampは11~dBAのセルフノイズを達成した。
In the dBA evaluation, PDS-Amp achieved a self-noise of 11~dBA.
Analysis using the SONY PCM-D50 and MATLAB showed that the self-noise of the dummy microphone with PDS-Amp was 11~dBA.
For reference, when calculated from the SNR ($>60$~dB) stated in the C9767 datasheet~\cite{CUI-C9767BB422LFP}, the conventional self-noise corresponds to approximately 34~dBA.
In a previous study by the authors~\cite{Obo-2025-HighPSRRLowOutputb}, the measured self-noise of the C9767 was reported as 23.1~dBA (under a different bias condition from the 1~G$\Omega$ condition, i.e., an unmodified C9767).
The self-noise for each condition is summarized as follows:
\begin{itemize}
\item PDS-Amp (custom photosensor, dummy microphone 12~pF): \textbf{11~dBA}
\item Estimate from C9767 datasheet SNR ($>60$~dB): approximately 34~dBA
\item Measured value in a prior study~\cite{Obo-2025-HighPSRRLowOutputb} (C9767, different bias condition): 23.1~dBA
\end{itemize}
Compared with any of these benchmarks, the noise reduction achieved by PDS-Amp is substantial.

\subsection{Test~3: Acoustic Recording Performance}

\begin{figure*}[t]
\centering
\includegraphics[width=177mm]{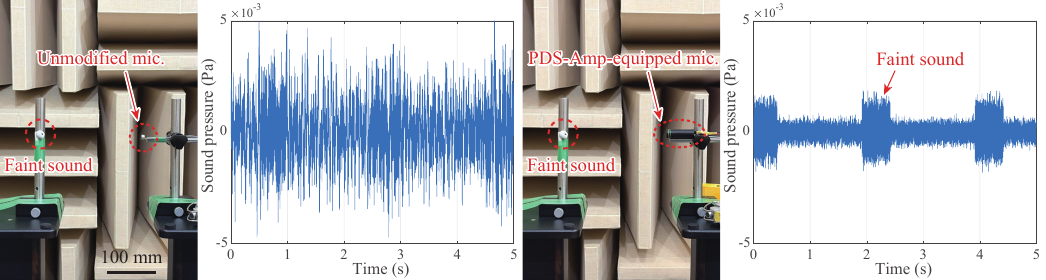}
\caption{Time-domain waveform comparison between the unmodified C9767 (left side) and the PDS-Amp-equipped C9767 (right side). A wireless earphone placed 200~mm from the microphone reproduced a 0.5-s vowel sound followed by 1.5-s silence (2-s cycle) in an anechoic chamber. Each microphone was individually calibrated with an acoustic calibrator (94~dB~SPL, 1~kHz). The sound pressure axis is identical for both panels. The PDS-Amp-equipped microphone shows a significantly reduced background self-noise, and the faint sound waveform is clearly visible.}
\label{fig_TimeDomain}
\end{figure*}

% 実ECMを用いた録音実験により、PDS-Ampのセルフノイズ低減効果を定性的に確認した。
The self-noise reduction effect of PDS-Amp was qualitatively confirmed through a recording experiment using an actual ECM.
Fig.~\ref{fig_TimeDomain} shows a comparison of time-domain waveforms between the unmodified C9767 (using the built-in JFET and built-in high-resistance element as-is) and the PDS-Amp-equipped C9767 (with the built-in JFET and built-in high-resistance element removed and replaced by the external PDS-Amp preamplifier).
The experiment was conducted in an anechoic chamber.
A wireless earphone placed 200~mm from the microphone reproduced a 0.5-s vowel sound followed by 1.5-s silence (2-s cycle).
The two microphones were recorded in separate trials under the same source and playback conditions; each microphone was individually calibrated using an acoustic calibrator (94~dB~SPL, 1~kHz), and the vertical axis is expressed in calibrated sound pressure (Pa) with an identical scale for both panels.
In the unmodified C9767, the faint sound is buried in background self-noise and is indiscernible in the waveform; however, auditory inspection of the recording confirms that the sound is indeed captured.
In contrast, in the PDS-Amp-equipped C9767, the background self-noise is significantly reduced, and the faint sound waveform is clearly visible.

%%%
% TimeDomain Figure

%%%%%%%%%%%%%%%%%%%%%%%%%%%%%%%%%%%%%%%%%%%%%%%%%%%%%
%% Section 5: Discussion
%%%%%%%%%%%%%%%%%%%%%%%%%%%%%%%%%%%%%%%%%%%%%%%%%%%%%
\section{Discussion}

% PDS-Ampの実験結果は、Section~IIで示した理論的予測と整合する。
The experimental results of PDS-Amp are consistent with the theoretical predictions presented in Section~II.
Section~II predicted that the thermal noise of the gate-bias resistor is the dominant factor of ECM self-noise and that noise reduction is achievable by replacing the bias element with a high-impedance, low-noise alternative.
The measured NSD comparison shown in Fig.~\ref{fig_CompPSD} demonstrates that PDS-Amp reduces the noise spectral density across the entire audible band, corroborating this theoretical prediction.
The particularly pronounced improvement in the low-frequency region indicates that the constraint of single-time-constant noise spectrum shaping by the conventional $R_{\mathrm{m}}C_{\mathrm{m}}$ circuit has been lifted.

% PDS-Ampで達成された11~dBAは、大型ダイヤフラムマイクロフォンの性能に匹敵する水準である。
The 11~dBA achieved by PDS-Amp is a level comparable to that of large-diaphragm microphones.
Table~\ref{tab_comparison} shows the self-noise of commercially available ultra-low-noise microphones (excluding shotgun microphones).
Among large-diaphragm condenser microphones (LDCs), the lowest-noise products achieve 3.7--5~dBA, but they require large diaphragms exceeding one inch and high-voltage biasing.
Among small-diaphragm condenser microphones (SDCs), the lowest-noise products are in the 10--15~dBA range, and measurement microphones are distributed in the 5.5--15~dBA range.
The 11~dBA achieved by PDS-Amp was obtained using a general-purpose 9~mm ECM
capsule costing approximately \$0.16, in contrast to the dedicated large-diaphragm
or studio-grade capsules employed in the reference microphones listed in
Table~\ref{tab_comparison}.
As shown in Fig.~\ref{fig_NFdBA}, the bulk of the self-noise distribution for commercially available ECMs is around 30~dBA.
The 11~dBA of PDS-Amp is well below this bulk zone and falls into the highest-performance category, comparable to LDCs costing several thousand dollars.
It is particularly noteworthy that this result was achieved with the C9767 (unit price approximately \$0.16 USD), an inexpensive general-purpose $\phi$9~mm ECM capsule, and the preamplifier circuit alone.
If a higher-performance ECM capsule than the C9767 were used, further noise reduction could be expected.
In other words, PDS-Amp is a circuit technique that achieves top-class self-noise performance even with an inexpensive ECM, without requiring a larger form factor or higher voltages.

%%%
% Comparison Table
\begin{table}[t]
\centering
\caption{Self-noise comparison between commercially available ultra-low-noise microphones and PDS-Amp (shotgun microphones excluded).}
\label{tab_comparison}
\begin{tabular}{llcc}
\hline
\textbf{Model} & \textbf{Type} & \textbf{Self-Noise} & \textbf{Price} \\
& & (dBA) & (USD) \\
\hline
Shure KSM44A & LDC & 4~\cite{Shure-KSM44A} & \$1,099 \\
Lewitt LCT 540 S & LDC & 4~\cite{Lewitt-LCT540S} & \$799 \\
Rode NT1 & LDC & 4.5~\cite{Rode-NT1} & \$154 \\
Audio-Technica AT5040 & LDC & 5~\cite{AudioTechnica-AT5040} & \$3,279 \\
B\&K Type 4955 & Meas. & 5.5~\cite{BK-Type4955} & --- \\
Sennheiser MKH 8020 & SDC & 10~\cite{Sennheiser-MKH8020} & \$1,499 \\
Neumann U 87 Ai & LDC & 12~\cite{Neumann-U87Ai} & \$3,200 \\
Sennheiser MKH 50-P48 & SDC & 12~\cite{Sennheiser-MKH50P48} & \$1,499 \\
B\&K Type 4188 & Meas. & 14.2~\cite{BK-Type4188} & --- \\
DPA 4006 & SDC & 15~\cite{DPA-4006} & \$2,960 \\
\hline
\textbf{PDS-Amp (this work)} & \textbf{ECM} & \textbf{11} & --- \\
\hline
\end{tabular}
\end{table}

% PDS-Ampは、大型化や高電圧化を伴わず、回路技術のみでセルフノイズを低減する手段である。
PDS-Amp is a means of reducing self-noise solely through circuit technique, without requiring a larger form factor or higher voltages.
Conventionally, self-noise reduction required large-aperture diaphragms or high-voltage biasing, making it difficult to simultaneously achieve miniaturization and low noise.
PDS-Amp reduces self-noise without constraints on diaphragm size or bias voltage through the replacement of the bias element and the introduction of a DC servo loop.
This feature enables retrofit application to miniature microphones and existing ECM capsules, improving noise performance without altering the conventional sensor design.

% なお、Test~1およびTest~2で得られた結果は、MLCC疑似マイクを用いた回路ノイズの評価であり、プリアンプ回路が達成しうるノイズフロアの下限を示すものである。
The results obtained in Test~1 and Test~2 represent circuit-noise evaluations using the MLCC dummy microphone and indicate the lower bound of the noise floor achievable by the preamplifier circuit.
When applied to an actual ECM capsule, the effective self-noise may increase beyond the dummy microphone value because of the addition of mechanical Brownian-motion noise of the diaphragm~\cite{Kim-2015-AcousticalThermalNoiseCapacitiveMEMS} and parasitic capacitance inside the capsule.
However, the DC insulation resistance of the C9767 diaphragm alone has been confirmed to be at least several T$\Omega$, and the noise current attributable to the diaphragm insulation resistance is sufficiently smaller than the input bias current of the preamplifier.
In general, accurately measuring the absolute value of the noise floor at this level presents difficulties.
Multiple factors are involved, including the ambient noise of the measurement environment, the accuracy of the calibrator, and the quantization noise of the recorder, so careful attention is required for precise evaluation.

% PDS-Ampの実用化に向けては、いくつかの課題が残されている。
Several challenges remain toward the practical deployment of PDS-Amp.
First, verification of the reproducibility and long-term stability of the custom photosensor is necessary.
Although epoxy sealing and oxygen-removal treatment suppress the oxidation of the zinc plate, accumulation of aging data is needed regarding long-term UV-induced degradation of the epoxy and changes in the hermeticity of the sealant.
Second, the addition of the DC servo loop increases circuit complexity, component count, and cost.
Third, the need for close proximity between the photoelectric element and the LED imposes constraints on packaging design.
These challenges may be addressed in the future through single-chip integration or fusion with MEMS technology.

% PDS-Ampの原理は、コンデンサマイクロフォンに限らず、広く静電容量型センサに適用可能である。
The principle of PDS-Amp is applicable not only to condenser microphones but also broadly to capacitive sensors.
Accelerometers and pressure sensors share the same structure of detecting changes in capacitance as condenser microphones.
Because thermal noise in the bias circuit can also be a source of self-noise in these sensors, the PDS-Amp technique is directly applicable.
Furthermore, pyroelectric sensors used in occupancy detectors and gas sensors share the capacitive operating principle and are promising candidates for PDS-Amp application.
For example, a high-precision occupancy sensor can reduce false detections, and a high-sensitivity accelerometer can contribute to cost-effective structural health monitoring (SHM) of bridges and tunnels~\cite{Crognale-2024-DevelopingTestingHighPerformanceSHM}.

%%%%%%%%%%%%%%%%%%%%%%%%%%%%%%%%%%%%%%%%%%%%%%%%%%%%%
%% Section 6: Conclusion
%%%%%%%%%%%%%%%%%%%%%%%%%%%%%%%%%%%%%%%%%%%%%%%%%%%%%
\section{Conclusion}

% 本論文は、静電容量型センサのセルフノイズを低減する回路技術PDS-Ampを提案した。
This paper proposed PDS-Amp, a circuit technique for reducing the self-noise of capacitive sensors.
PDS-Amp replaces the gate-bias resistor---the dominant source of self-noise---with a photoelectric element and stabilizes the bias voltage through a DC servo loop.
In conventional resistor biasing, the noise low-pass cutoff frequency and the signal high-pass cutoff frequency were constrained to the same $R_{\mathrm{m}}C_{\mathrm{m}}$ time constant; PDS-Amp enables them to be designed independently.

% コンデンサマイクロフォンを対象とした実験により、PDS-Ampの有効性を実証した。
The effectiveness of PDS-Amp was demonstrated through experiments on condenser microphones.
In the quantitative evaluation using a dummy microphone (MLCC 12~pF), a self-noise of 11~dBA was achieved.
In addition, a recording experiment using an actual ECM (C9767) qualitatively confirmed that the self-noise is substantially reduced compared with the unmodified product, enabling the faint sound waveform to be clearly resolved in the time domain.

% PDS-Ampは、マイクロフォンに限らず、静電容量型センサ全般の性能向上に貢献する基盤技術である。
PDS-Amp is a foundational technology that contributes to performance improvement not only for microphones but for capacitive sensors in general.
Applications to accelerometers, pressure sensors, pyroelectric sensors, and other sensors sharing the same operating principle are anticipated.
Future work will focus on advancing the practical deployment of PDS-Amp through improved stability of the photoelectric element and single-chip integration.

%%%%%%%%%%%%%%%%%%%%%%%%%%%

%%%%%%%% 
\bibliographystyle{IEEEtran}
\bibliography{library}
%%%%%%%%

\begin{IEEEbiography}
[{\includegraphics[width=1in,height=1.25in,clip,keepaspectratio]{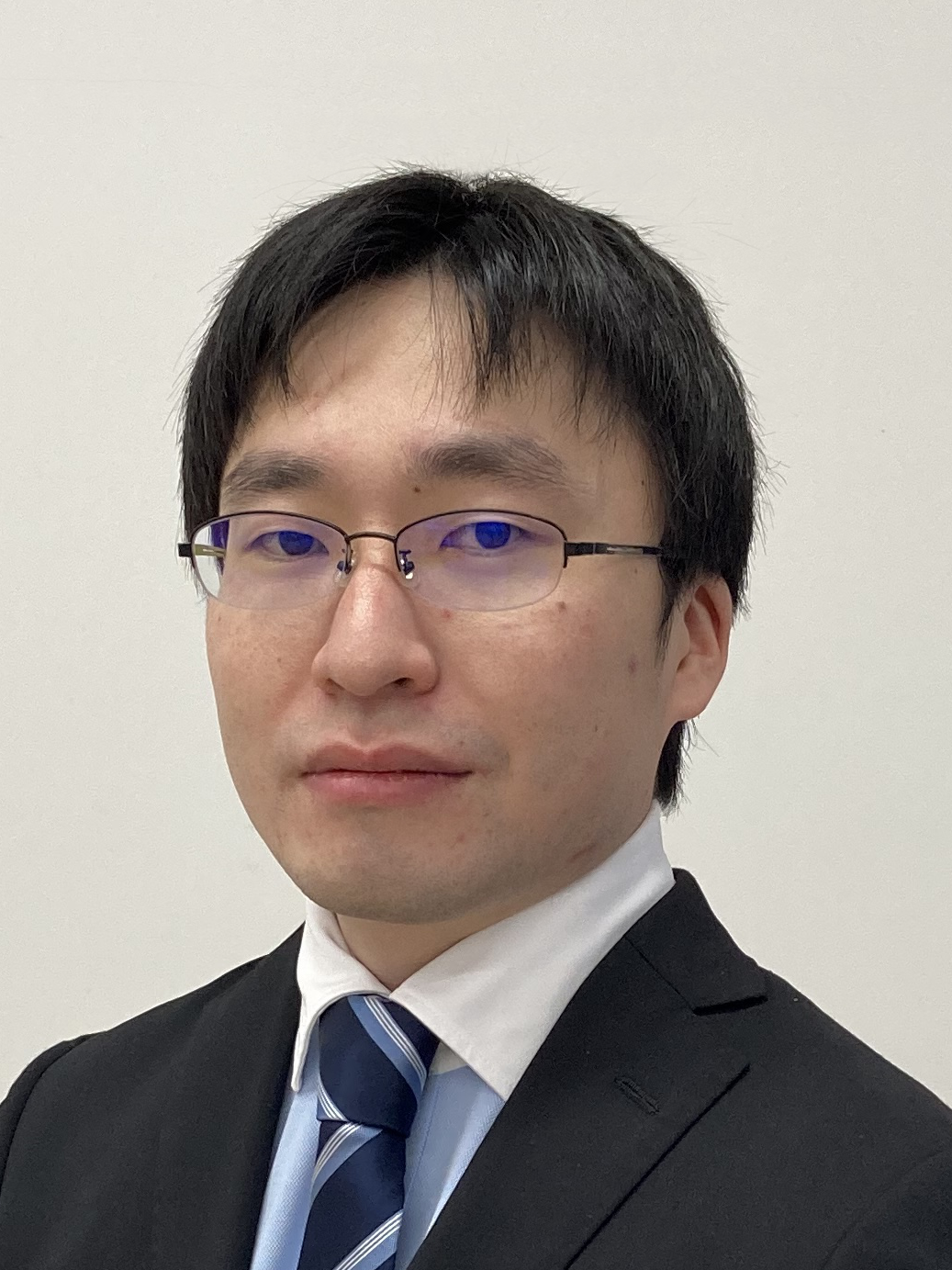}}]{Hirotaka Obo}
~received his Ph.D. degree from the University of Tsukuba, Tsukuba, Japan, in 2024.
His research interests include sound source localization, ultrasound electronics, electronic circuits, agricultural information engineering, and instrumentation engineering.
From April 2021 to September 2023, he was a research fellow of the Japan Society for the Promotion of Science (DC2). He is currently a researcher at the National Agriculture and Food Research Organization (NARO) and also a visiting researcher at the University of Tsukuba.
He is a member of the Acoustical Society of Japan (ASJ), the Society of Agricultural Structures, Japan (SASJ), and the Institute of Electronics, Information and Communication Engineers (IEICE).
\end{IEEEbiography}

\begin{IEEEbiography}
[{\includegraphics[width=1in,height=1.25in,clip,keepaspectratio]{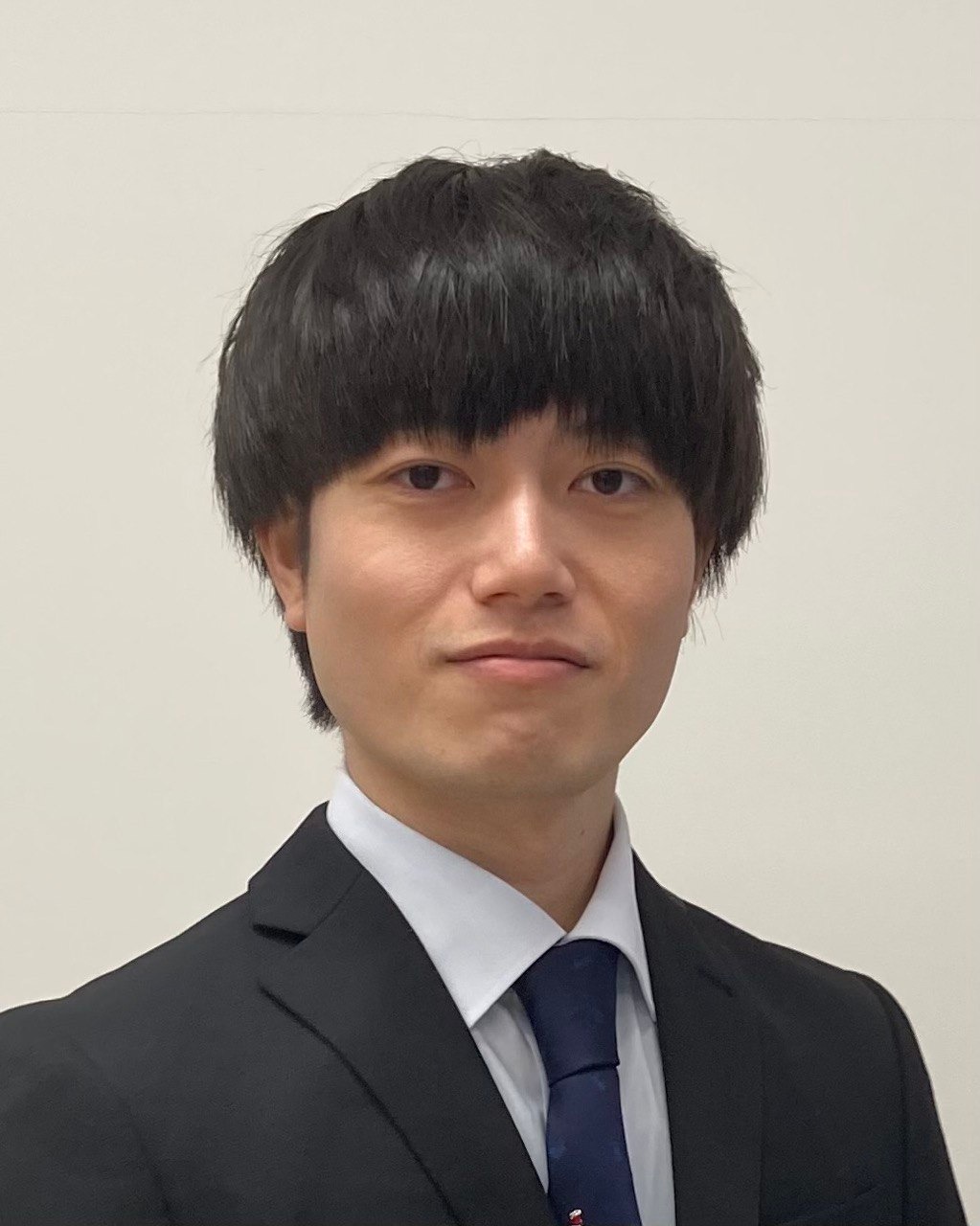}}]{Atsushi Tsuchiya}
received his Ph.D. from the University of Tsukuba, Japan, in 2025.
Since April 2022, he has been a Research Fellow of the Japan Society for the Promotion of Science (DC1).
Since 2025, he has been a researcher at the University of Tsukuba, Japan.
His research interests include self-localization using acoustic echo, ultrasound electronics, underwater acoustics, mobile robotics engineering, and civil engineering.
He is a member of the IEEE, the
Institute of Electronics; Information and Communication Engineers (IEICE));, and the Japan Society of Civil Engineering (JSCE).

\end{IEEEbiography}

\begin{IEEEbiography}
[{\includegraphics[width=1in,height=1.25in,clip,keepaspectratio]{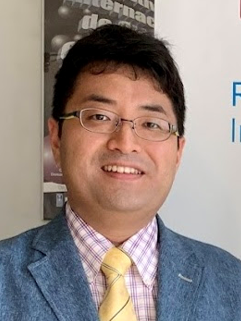}}]{Tadashi Ebihara}
~(Senior Member, IEEE)~received his Ph.D. degree from the University of Tsukuba, Tsukuba, Japan, in 2010. From September 2013 to December 2013, he was a visiting professor with the Delft University of Technology, The Netherlands. He is currently an associate professor with the Faculty of Engineering, Information and Systems, University of Tsukuba. His research interests include mobile communications and their applications to underwater acoustic communication systems. He received the Research Fellowship for Young Scientists (DC1) from the Japan Society for the Promotion of Science (JSPS) for the years 2009 and 2010. He received the 2017 IEEE Oceanic Engineering Society Japan Chapter Young Researcher Award.
\end{IEEEbiography}

\begin{IEEEbiography}
[{\includegraphics[width=1in,height=1.25in,clip,keepaspectratio]{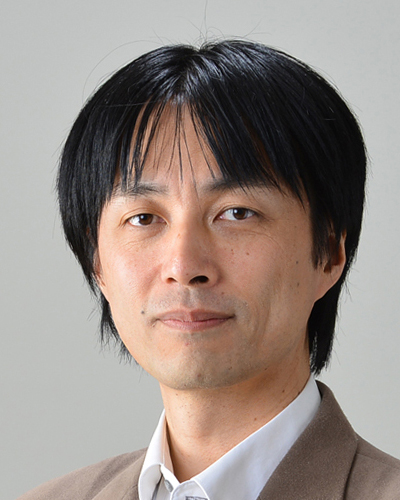}}]{Naoto Wakatsuki}
~received his B.Eng., M.Eng., and D.Eng. degrees from the University of Tsukuba in 1993, 1995, and 2004, respectively. He was with Okayama University, from 1995 to 2001, and with Akita Prefectural University, from 2001 to 2006. He is currently a full professor with the Faculty of Engineering, Information and Systems, University of Tsukuba. His research interests include acoustic instrumentation, simulation-based visualization, vibration sensors and actuators, acoustical engineering, musical acoustics, and inverse problems. Dr. Wakatsuki is affiliated with the Acoustical Society of Japan, the Acoustical Society of America, the Society of Agricultural Structures, and the Japan Society for Simulation Technology.
\end{IEEEbiography}

\end{document}